\documentclass[pra,a4paper,aps,twocolumn,showpacs,superscriptaddress,groupedaddress]{revtex4}
\usepackage[dvips]{graphicx}
\usepackage{ae}
\usepackage[T1]{fontenc}
\usepackage[ansinew]{inputenc}
\usepackage{amsmath}
\usepackage{amssymb}
\usepackage{graphicx}
\usepackage{caption}
\usepackage{color}
\usepackage[colorlinks]{hyperref}
\usepackage{lscape}
\begin{document}
	\title{Detection of Genuine Multipartite Entanglement in Quantum Network Scenario}

\author{Biswajit Paul}
\email{biswajitpaul4@gmail.com}
\affiliation{Department of Mathematics, South Malda College, Malda, West Bengal, India}

\author{Kaushiki Mukherjee}
\email{kaushiki_mukherjee@rediffmail.com}
\affiliation{Department of Mathematics, Government Girls’ General Degree College, Ekbalpore, Kolkata, India.}

\author{Sumana Karmakar}
\email{sumanakarmakar88@gmail.com}
\affiliation{Department of Applied Mathematics, University of Calcutta, 92, A.P.C. Road, Kolkata-700009, India.}

\author{Debasis  Sarkar}
\email{dsarkar1x@gmail.com,dsappmath@caluniv.ac.in}
\affiliation{Department of Applied Mathematics, University of Calcutta, 92, A.P.C. Road, Kolkata-700009, India.}

\author{Amit Mukherjee}
\email{amitisiphys@gmail.com}
\affiliation{Physics and Applied Mathematics Unit, Indian Statistical Institute, 203, B. T. Road, Kolkata 700108 , India.}

\author{Arup Roy}
\email{arup145.roy@gmail.com}
\affiliation{Physics and Applied Mathematics Unit, Indian Statistical Institute, 203, B. T. Road, Kolkata 700108 , India.}

\author{Some Sankar Bhattacharya}
\email{somesankar@gmail.com}
\affiliation{Physics and Applied Mathematics Unit, Indian Statistical Institute, 203, B. T. Road, Kolkata 700108 , India.}

\begin{abstract}
Experimental demonstration of entanglement needs to have a precise control of experimentalist over the system on which the measurements are performed as prescribed by an appropriate entanglement witness.  To avoid such trust problem, recently device-independent entanglement witnesses (\emph{DIEW}s) for genuine tripartite entanglement have been proposed where witnesses  are capable of testing genuine entanglement without precise description of Hilbert space dimension and measured operators i.e apparatus are treated as black boxes. Here we design a protocol for enhancing the possibility of identifying genuine tripartite entanglement in a device independent manner. We consider three mixed tripartite quantum states none of whose genuine entanglement can be detected by applying standard \emph{DIEW}s, but their genuine tripartite entanglement can be detected by applying the same when distributed in some suitable entanglement swapping network.
\end{abstract}

\maketitle

\section{I. INTRODUCTION}
Entanglement is one of the most intriguing and most fundamentally non-classical phenomena in quantum physics. A bipartite quantum state without entanglement is called separable. A multipartite quantum state that is not separable with respect to any bi-partition is said to be genuinely multipartite entangled \cite{Guh}. This type of entanglement is important  not only for research concerning the foundations of quantum theory but also in quantum information protocols and quantum tasks such as extreme spin squeezing \cite{Sor}, high sensitivity in some general metrology tasks \cite{Hyl}, quantum computing using cluster states \cite{Rau}, measurement-based quantum computation \cite{Bri} and multiparty quantum network \cite{Mur,Hil,Sca,Zha}. Several experiments have been conducted so far for generation of genuine multipartite entanglement \cite{Yao,Gao,Mon}. However, detection of this kind of resource in an experiment turns out to be quite difficult. Experimental demonstration of genuine multipartite entanglement is generally performed with one of the two following techniques: tomography of the full quantum state \cite{Alt,Lvo}, or evaluation of an entanglement witness \cite{Guh}. But both of these techniques face some common drawbacks viz. requirement of precise control(by the experimentalist) over the system subjected to measurements and sensitivity of these techniques to systematic errors \cite{Ros}. \\
However, there exists a way to avoid this sort of drawbacks. Such an alternative method is provided by using some specific Bell-type inequalities \cite{Bel}.  Bell inequality was first designed by John Bell to explain the incompatibility of quantum predictions with local-realism\cite{Bel}. Till date various types of Bell inequalities have been designed for the purpose of detection of nonlocality of correlations where any precise control of the device by the experimentalist is not needed. Now presence of entanglement is a necessary resource for generation of nonlocal correlations. In this context some specific type of Bell inequalities have been proposed to detect entanglement, more specifically genuine multipartite entanglement(GME) certified from statistical data only. To be precise, if the value of a Bell expression in multipartite scenario exceeds the value obtained due to measurements on biseparable quantum states, then the presence of genuine entanglement can be guaranteed. This technique to detect genuine multipartite entanglement in device-independent manner was first introduced in \cite{See,Nag,Uff,Sev} followed by an extensive formalization by Bancal et al. \cite{Ban}. In particular they introduced the term Device-Independent Entanglement Witness(DIEW) of genuine multipartite entanglement for such Bell expressions. Later, Pal \cite{Pal} and Liang et al. \cite{Lia} developed other DIEWs for detecting genuine multipartite entanglement. Throughout the paper, we refer to the procedure of detecting genuine entanglement as device-independent entanglement detection(DIED) and the entanglement detected in device-independent way as device-independent entanglement(DIE).

Entanglement swapping\cite{Zuk} as a resource has been used in a number of quantum information processing tasks such as entanglement concentration or distillation, purification, speeding up the distribution of entanglement, correction of amplitude errors developed due to propagation, activation of nonlocality etc. Entanglement swapping in tripartite scenario mainly involves a network of three parties say, Alice, Bob and Charlie. The procedure of entanglement swapping was first generalized for multi-party scenario in \cite{Bos}. In the present paper, we address the following questions: consider some tripartite states whose genuine entanglement cannot be detected by applying some standard DIEWs \cite{See,Uff,Ban,Lia}, now is it possible to find some suitable entanglement swapping process, after which the genuine entanglement of swapped state can be detected by those DIEWs? We answer this question affirmatively and have designed a protocol based on entanglement swapping procedure by which genuine entanglement of the tripartite state resulting from multiple swapping can be detected in a device independent way , i.e. without any reference of the device involved. Precisely speaking, this new protocol enhances the regime of DIED for tripartite quantum states. In this context another important question is whether one can enhance detection of genuine entanglement in a semi-device independent way(corresponding to phenomenon of quantum steering). We also answer this question affirmatively.

The rest of this paper has been organized as follows: section II deals with some mathematical preliminaries and a brief overview of some standard DIEWs. In section III we design the protocol involving entanglement swapping procedure followed by detailed discussion on enhancement of entanglement detection by using some standard DIEWs in section IV. In section.V we have used this protocol to enhance genuine entanglement in a semi device independent way. Finally we conclude with a brief discussion regarding the importance of this work and possible further extensions.

\section{BACKGROUND}
\subsection{Notion of DIEWs}
Violation of Bell inequality by quantum mechanical systems always indicates presence of entanglement. Thus a Bell inequality can be considered as a suitable candidate for detecting the presence of entanglement in a device independent way unlike the standard procedures like state tomography or use of entanglement witnesses where experimentalist needs to trust the experimental apparatus. This is because detection of entanglement using Bell inequality solely depends on the statistical data. To characterize genuine entanglement in a device-independent way for tripartite scenario where each of the three subsystems, one of m possible measurements can be performed, yielding one of two possible outcomes. The measurement settings are denoted by $x$, $y$, $z$ $\in$ $\{0,1,2,...m-1\}$ and their outputs by $a$, $b$, $c$ $\in$ \{-1,1\} for Alice, Bob and Charlie respectively. The experiment is thus characterized by the joint probability distribution $p(a b c|x y z)$. The correlations $P(abc|xyz)$ can be categorized as bi-separable if they can be reproduced through the measurements on a tripartite bi-separable state $\rho_{bi}$ where $\rho_{bi} =$
 \begin{equation}\label{e1}
 \sum_{\lambda} p_{\lambda} \rho_{\lambda} ^{A} \bigotimes \rho_{\lambda} ^{BC} + \sum_{\mu} p_{\mu} \rho_{\mu} ^{B} \bigotimes \rho_{\mu} ^{AC} + \sum_{\nu} p_{\nu} \rho_{\nu} ^{C} \bigotimes \rho_{\nu} ^{AB}.
 \end{equation}
 Here $0\,\leq\,p_{\lambda},\,p_{\mu},p_{\nu}\,\leq\,1$ and $\sum_{\lambda}p_{\lambda}+\sum_{\mu}p_{\mu}+\sum_{\nu}p_{\nu}=1.$
 To be precise, if there exists a state of the form given by Eq.(\ref{e1}) in some Hilbert space $\mathcal{H}$ and some suitable local measurement operators $M_{a|x}$, $M_{b|y}$ and $M_{c|z}$(without loss of generality these operators can be considered to be projection operators satisfying the restriction $M_{a|x} M_{a^{'}|x}=\delta_{a,a^{'}}M_{a|x}$ and $\sum_{a}M_{a|x}=\textbf{I}$) such that:
 \begin{equation}\label{e2}
    p(a b c|x y z) = tr[M_{a|x}\bigotimes M_{b|y}\bigotimes M_{c|z} \rho_{bi}]
 \end{equation}
 If the correlations are not biseparable, then the state used is surely a genuine tripartite entangled state. Such a conclusion can be drawn independent of the corresponding Hilbert space dimension. Equivalently, biseparable quantum correlations can also be decomposed as,\\
 $$P(abc|xyz) =\sum_{k}P_{Q}^{k}(ab|xy ) P_{Q}^{k}(c|z)+$$
 \begin{equation}\label{e3}
   \sum_{k} P_{Q}^{k}(ac|xz) P_{Q}^{k}(b|y )+ \sum_{k} P_{Q}^{k}(bc|yz) P_{Q}^{k}(a|x)
\end{equation}
where $P^k_Q(ab|xy)$ and $P^k_Q(c|z)$ denote arbitrary two party and one party quantum correlations respectively. So they are of the form: $P^k_Q(ab|xy)=tr[M_{a|x}^k{a|x}\bigotimes M_{b|y}^k\rho^k_{AB}]$ and $P^k_Q(c|z)=tr[M_{c|z}^k\rho^k_{C}]$ for some unnormalized quantum states $\rho^k_{AB}$, $\rho^k_{C}$ and measurement operators $M_{a|b}^k,M_{b|y}^k, M_{c|z}^k$. \\
Let $Q_3$ denotes the set of tripartite quantum correlations and $Q_{2|1}$ denotes the set of biseparable quantum correlations. Clearly, $ Q_{2|1}\subseteq Q_3.$ The set $Q_{2|1}$ being convex, can be characterized by linear inequalities. DIEWs of genuine tripartite entanglement correspond to those inequalities(Bell inequalities) that separate the sets of $Q_3$ and $Q_{2|1}$. Now as $Q_{2|1}$ has infinite number of extremal points so there exist many such DIEWs separating genuine entanglement from bi-separable entanglement. In recent times many such DIEWs are designed for detecting genuine tripartite entanglement in a device independent way \cite{See,Nag,Uff,Sev,Ban,Pal,Lia}. As already mentioned in the introduction, Bancal \cite{Ban} was the first to formalize the concept of device independent detection of entanglement introducing the term DIEW for detecting genuine multipartite entanglement. In this context, one can consider the DIEW provided by the Mermin polynomial \cite{Mer} as the most simple example for detecting genuine tripartite entanglement \cite{See}.
In \cite{Uff} Uffink, designed another non linear Bell-type inequality which has been extensively used for this purpose.
In recent times, Bancal et al. gave more efficient $3$-settings Bell inequality which can be used as a DIEW to detect genuine tripartite entanglement \cite{Ban}.
More than $3$ setting DIEWs are also provided in \cite{Pal}. However in our present topic of discussion, we restrict our search for not more than $3$-settings Bell inequalities due to obvious computational complexity.  More recently another $2$ settings DIEW was designed by Liang et al.\cite{Lia}(see Appendix.\ref{AppA}).
As detection of  genuine nonlocality by any Bell inequality implies genuine entanglement \cite{Sve,Bac} so it is a DIEW for detecting genuine entanglement. However the converse is not necessarily true. After discussing about DIEW and their advantages over usual procedures of detecting entanglement experimentally, we are now in a position to use them for our purpose of detecting genuine entanglement in an entanglement swapping protocol. This in turn helps to enhance the chance of genuine tripartite entanglement being detected in a device independent way. But before that we illustrate our multiple entanglement swapping scenario.

\section{Multiple Entanglement Swapping Procedure}\label{swap}

\begin{figure}[b!]
	\includegraphics[scale=0.4]{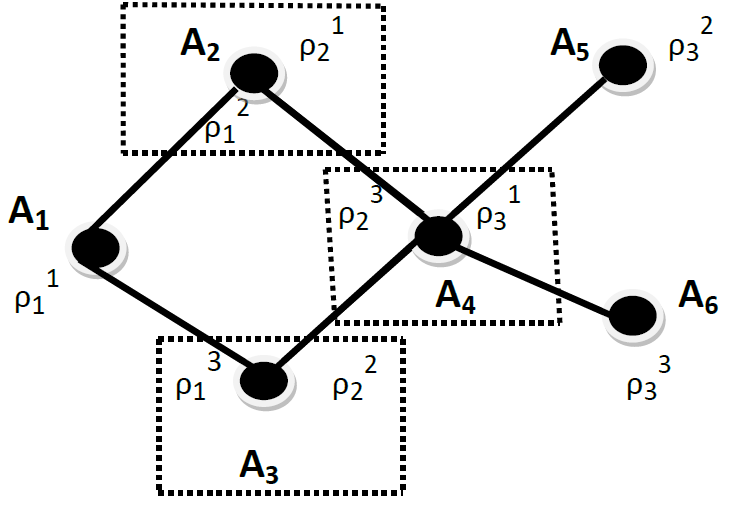}
	\caption{Swapping scheme}\label{sets}
\end{figure}
Consider the multiple entanglement swapping network given in Fig.\ref{sets}. It is a network of six space-like separated observers. Three tripartite quantum states $\rho_i(i=1,2,3)$ are used in the network. State $\rho_1$ is shared among the parties $A_i(i=1,2,3)$ such that $j^{th}$ particle($\rho_1^j$) of $\rho_1$ is with party $A_j(j=1,2,3)$ respectively. State $\rho_2$ is shared among $A_2$, $A_3$ and $A_4$ with the specification that $j^{th}$ qubit($\rho_2^j$) is sent to party $A_{j+1}(j=1,2,3)$. The remaining state $\rho_3$ is shared among $A_4$, $A_5$ and $A_6$ such that party $A_{j+3}$ holds $j^{th}(j=1,2,3)$ particle of $\rho_3$($\rho_3^j$). So each of the three parties $A_2$, $A_3$ and $A_4$ holds two particles: $A_2$ holds $\rho_1^2$ and $\rho_2^1$; $A_3$ holds $\rho_1^3$ and $\rho_2^2$; $A_4$ holds $\rho_2^3$ and $\rho_3^1$. Now in the preparation stage, each of the three parties $A_i(i=1,2,3)$  performs Bell basis measurements on two of the three particles that each of them holds: $A_1$ performs Bell basis measurement on $3^{rd}$ particle of $\rho_2$($\rho_2^3$) and $1^{st}$ particle of $\rho_3$($ \rho_3^1$); $A_2$ performs Bell basis measurement on $2^{nd}$ particle of $\rho_1$($\rho_1^2$) and $1^{st}$ particle of $\rho_2$($ \rho_2^1$); $A_3$ performs Bell basis measurement on $3^{rd}$ particle of $\rho_1$($\rho_1^3$) and $2^{nd}$ particle of $\rho_2$($ \rho_2^2$). After all the three parties have performed Bell basis measurement on their respective particles, they communicate the results among themselves, as a result of which $\rho_4$ is generated at the end of the preparation stage. Clearly $\rho_4$ varies with the output of the Bell measurements. The final state $\rho_4$ is obtained from the initial states $\rho_i (i=1,2,3)$ by means of post-selecting on particular results of local measurements, in particular Bell basis measurements performed on these states($\rho_i (i=1,2,3)$). For example, let us consider the case when all the parties obtain outputs corresponding to $|\psi^{\pm}\rangle=\frac{|01\rangle\pm|10\rangle}{\sqrt{2}}$. If the output of all measurements correspond to $|\psi^{+}\rangle$($|\psi^{-}\rangle$), the resultant state $\rho_4^{+}$($\rho_4^{-}$) is given by:
\begin{eqnarray}
\rho_4^{\pm}=&& \langle \psi^{\pm}|^{A_2}\otimes\langle\psi^{\pm}|^{A_3}\otimes\langle \psi^{\pm}|^{A_4}\nonumber\\&&(\rho_1\otimes\rho_2\otimes\rho_3)\nonumber\\ &&|\psi^{\pm}\rangle^{A_2}\otimes |\psi^{\pm}\rangle^{A_3}\otimes|\psi^{\pm}\rangle^{A_4}
\end{eqnarray}

So preparation stage of this protocol can be considered as a particular instance of Stochastic Local Operation and Classical Communication (SLOCC). After $\rho_4^{\pm}$ is generated and shared among the parties in the preparation stage, each of the three parties $A_1$, $A_5$ and $A_6$ performs projective measurement on the state $\rho_4^{\pm}$ in the measurement stage. Now if the correlations generated from $\rho_4^{\pm}$ exhibit violation of any DIEW under the context that the initial  states $\rho_i(i=1,2,3)$ fail to reveal the same, then that guarantees enhancement of DIED in our protocol.

\section{ENHANCEMENT OF DEVICE-INDEPENDENT ENTANGLEMENT DETECTION possibility}
In this section we deal with the procedure of enhancing DIED of tripartite quantum states in terms of expanding the set of states by using the multiple entanglement swapping protocol described in Fig.\ref{sets}. For this we provide an explicit example. Initially we consider three tripartite quantum states $\rho_i$$(i=1,2,3)$ with some restricted range of state parameters, for each of which none of the DIEWs proposed in the literature\cite{Mer,Lia,Ban} can detect genuine entanglement. These states, after being used in the multiple entanglement swapping network(Fig.\ref{sets}), generates a state $\rho_4$ whose genuine entanglement can be detected in a device-independent manner.\\
 Let the three initial states be given by:
 \begin{equation}\label{1}
 \rho_1 = p |\psi_f\rangle\langle \psi_f|+(1-
 p)|001\rangle\langle001|
 \end{equation}
 with $|\psi_f\rangle=\cos\theta|000\rangle+\sin\theta|111\rangle$, $0\leq\theta\leq \frac{\pi}{4}$ and $0\leq p\leq 1$;
 \begin{equation}\label{2}
 \rho_2=  p_1 |\psi_m^{+}\rangle\langle \psi_m^{+}|+(1-p_1)|010\rangle\langle010|
 \end{equation}
 with $|\psi_m^{+}\rangle=\frac{|000\rangle+|111\rangle}{\sqrt{2}}$ and $0\leq p_1\leq 1$;
 \begin{equation}\label{3}
 \rho_3=  p |\psi_l\rangle\langle \psi_l|+(1-p)|100\rangle\langle100|
 \end{equation}
 with $|\psi_l\rangle=\sin\theta|000\rangle+\cos\theta|111\rangle$. Now each of the three parties $A_2$, $A_3$ and $A_4$ performs Bell  basis measurement on their respective particle. As already stated before, the output state depends on the outputs of the Bell measurements performed. For instance when  $|\psi^{\pm}\rangle=\frac{|01\rangle\pm|10\rangle}{\sqrt{2}}$ is obtained as the output odd number of times, a resultant state $\rho_4^{\pm}$ is obtained which after correcting phase term is given by:
 \begin{equation}\label{4}
 \rho_4^{\pm} = p_f |\psi_m^{\pm}\rangle\langle \psi_m^{\pm}| + (1-p_f)|100\rangle\langle100|
 \end{equation}
 where $|\psi_m^{\pm}\rangle=\frac{|000\rangle\pm |111\rangle}{\sqrt{2}}$ and $p_f= \frac{2 p \cos^{2}\theta}{1+p \cos2\theta}$. Clearly $\rho_4^{\pm}$ is independent of $p_1$, but the probability of obtaining $\rho_4^{\pm}$ directly depends on it. Here reader must note that $\rho_4^{\pm}$ can also be generated for some other combination of swapping networks together with some different arrangement of particles in between the parties $A_i(1,...,6)$ and for different outputs of the Bell measurement.
 To detect DIE of each of the states $\rho_i(i=1,2,3,4)$ we obtain the condition for which they violate each of the DIEWs(see Appendix.\ref{AppA}) given in \cite{See,Uff,Ban,Lia}. Among all, the $3$-settings Bell inequality
$$\frac{\sqrt{3}}{2}(\langle A_0B_0C_0\rangle - \langle A_2B_0C_0\rangle -\langle A_1B_1C_0\rangle - \langle A_2B_1C_0\rangle$$
$$- \langle A_0B_2C_0\rangle -\langle A_1B_2C_0\rangle-\langle A_1B_0C_1\rangle-\langle A_2B_0C_1\rangle$$
$$- \langle A_0B_1C_1\rangle-\langle A_1B_1C_1\rangle-\langle A_0B_2C_1\rangle$$
$$+\langle A_2B_2C_1\rangle-\langle A_0B_0C_2\rangle-\langle A_1B_0C_2\rangle-\langle A_0B_1C_2\rangle +$$
\begin{equation}\label{baneq}
\langle A_2B_1C_2\rangle + \langle A_1B_2C_2\rangle + \langle A_2B_2C_2\rangle) \leq 9
\end{equation}
given by Bancal et al.\cite{Ban} is the most efficient DIEW for each of $\rho_i(i=1,2,3,4)$(see Table.\ref{table}). Here $\langle A_\alpha B_\beta C_\gamma\rangle$ designate the expected value of the product of three $\pm 1$ observables, $A_\alpha$, $B_\beta$, $C_\gamma$.   \\

\begin{figure}
\includegraphics[scale=0.3]{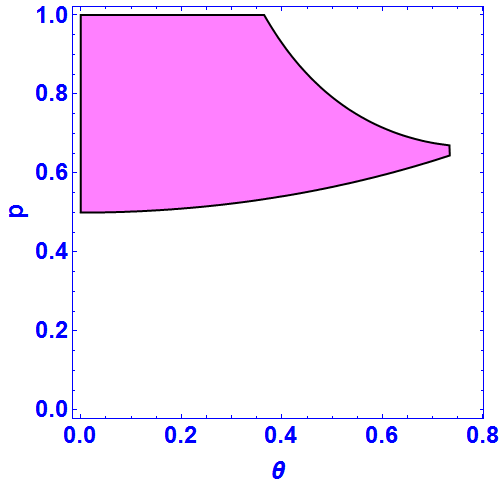}
\caption{(color online)\emph{The shaded region denotes the range of parameters of states $\rho_1$ and $\rho_3$( $p_1\leq \frac{2}{3}$) for which the $3$-settings Bell inequality (Eq.(\ref{baneq})) is violated only after distributing them in the multiple entanglement swapping network as described in Sec.\ref{swap}, i.e. neither of the three initial states violate Eq.(\ref{baneq}) whereas the final state violates it. Hence this region gives the range where DIED is enhanced.}}\label{pic2}
\end{figure}

Now the initial states $\rho_i$$(i = 1, 2, 3)$ do not violate Eq.(\ref{baneq}) if and only if(see Appendix.\ref{AppA})
$$p_1 \leq \frac{2}{3}$$  and
\begin{equation}\label{b1}
  p \leq \frac{2}{3 \sin 2 \theta}
\end{equation}
The condition of violation of Eq.(\ref{baneq}) for the final states $\rho_4^{\pm}$ is given by(see Appendix.\ref{AppA}):
\begin{equation}\label{b2}
  p > \frac{1}{\cos^2\theta + 1}
\end{equation}
There exists a range of the state parameters where the initial states $\rho_i(i=1,2,3)$ do not violate Bancal's 3-settings Bell inequality, but after distributing them in the multiple entanglement swapping network, final states $\rho_4^{\pm}$ violates it. The range of state parameters in which detection of DIE is enhanced by this entanglement swapping protocol(see Fig.\ref{pic2}) is given by: $p_1 \leq \frac{2}{3}$ and
\begin{equation}\label{r1}
  \frac{1}{\cos^2\theta + 1} < p \leq \frac{2}{3 \sin 2 \theta}.
\end{equation}
The restrictions imposed on the state parameters(Fig.\ref{pic2}) indicate that DIED is enhanced at the end of the swapping procedure.
In this context, it is interesting to note that the probability of success of this protocol $p^{succ}$ is given by $$p^{succ}=\frac{1}{2}pp_1[1+p \cos 2\theta]\sin^2\theta.$$ 
In recent times there has been experimental implementation of DIED \cite{Bar} and also experimental demonstration of  entanglement swapping \cite{Wpa,Goe}. So our procedure of enhancing DIED method can also be demonstrated experimentally within the scope of current technology.
The fact that this explicit example shows enhancement of the possibility of entanglement detection in a device independent manner indicates that for any genuinely entangled tripartite state($\rho$, say) it may be possible to design a suitable swapping protocol via which entanglement of the final state resulting from the protocol using many copies of the initial state($\rho$), can be detected even when the same cannot be detected for $\rho$ itself.

\begin{widetext}

 \begin{table}[htp]
	\centering
	\captionsetup{justification=centering}
\begin{tabular}{|c|c|c|c|c|}
			\hline
			DIEW &  Violation by $\rho_1$ &  Violation by $\rho_2$ & Violation by $\rho_3$&Enhanced range\\
			\hline
		Mermin\cite{Mer}& $p$$>$$\frac{1}{\sqrt{2}\sin2\theta}$ &  $p_1$$>$$\frac{1}{\sqrt{2}}$&$p$$>$$ \frac{1}{\sqrt{2}\sin2\theta } $,\, &$\frac{1}{(2\sqrt{2}-2) \cos^2 \theta + 1}$$<$$p$$\leq$$\frac{1}{\sqrt{2}\sin2\theta}$\\
			\hline
		Uffink\cite{Uff} &	$p$$>$$\frac{1}{\sqrt{2}\sin2\theta}$ &  $p_1$$>$$\frac{1}{\sqrt{2}}$&$p$$>$$\frac{1}{\sqrt{2}\sin2\theta }$,\, &$\frac{1}{(2\sqrt{2}-2) \cos^2 \theta + 1}$$<$$p$$\leq$$\frac{1}{\sqrt{2}\sin2\theta}$\\
			\hline
		Bancal et al.\cite{Ban} &	$p$$>$$\frac{2}{3\sin2\theta}$ &  $p_1$$>$$\frac{2}{3}$&$p$$>$$\frac{2}{3\sin2\theta}$,\,&$\frac{1}{\cos^2 \theta + 1}$$<$$p$$\leq$$\frac{2}{3\sin2\theta} $\\
			\hline
		Liang et al.\cite{Lia} &	$p$$>$$\frac{3\sqrt{2}}{5\sin2\theta}$ &  $p_1$$>$$\frac{3\sqrt{2}}{5}$&$p$$>$$\frac{3\sqrt{2}}{5\sin2\theta}$,\, &$\frac{3\sqrt{2}}{5\sin2\theta}$$<$$p$$\leq$$\frac{1}{(\frac{5\sqrt{2}}{3}-2)\cos^2\theta + 1}$\\
			\hline
\end{tabular}\\
\caption{ The condition of violation of each of the DIEWs given in \cite{Mer,Uff,Ban,Lia} for each of the states($\rho_i(i=1,2,3)$) are enlisted here. These conditions restrict the state parameters of the corresponding states such that these in turn give the enhanced region for detection of DIE in the swapping procedure. Moreover comparison of these restrictions(for each of these states) in turn clearly justifies our claim that the DIEW given by Bancal et al.\cite{Ban} emerges to be efficient tool for the detection of genuine entanglement in a device independent way.}\label{table}

\end{table}
\label{table1}
\end{widetext}

\section{ENHANCEMENT OF SEMI DEVICE-INDEPENDENT ENTANGLEMENT DETECTION POSSIBILITY}
Cavalcanti \emph{et.} al. in \cite{Cav} have provided an inequality which detect genuine entanglement in semi device independent way. The inequality looks like:
\begin{flalign}\label{semi}
 & 1-0.1831(\langle A_3 B_3 \rangle+ \langle A_3 Z \rangle+ \langle B_3 Z\rangle)-0.2582(\langle A_1 B_1 X \rangle&&\nonumber\\
 &-\langle A_1 B_2 Y \rangle- \langle A_2 B_1 Y\rangle-\langle A_2 B_2 X \rangle)\geq 0&&
\end{flalign}
 Following the same procedure as that for device independent case, it is observed that the initial states $\rho_i$$(i = 1, 2, 3)$ do not violate Cavalcanti et al. inequality if and only if\\
$$p \geq \frac{1.1831}{0.7324+1.0328 \sin[2\theta]}$$  and
\begin{equation}\label{b1}
  p_1\geq 0.670236
\end{equation}
Now, the condition of violation of the inequality(Eq.(\ref{semi})) for the final state $\rho_4$ is given by:
\begin{equation}\label{b2}
  p > \frac{2p \cos^2[\theta]}{1+p \cos[2 \theta]}
\end{equation}
\par
\begin{figure}
	\includegraphics[scale=0.35]{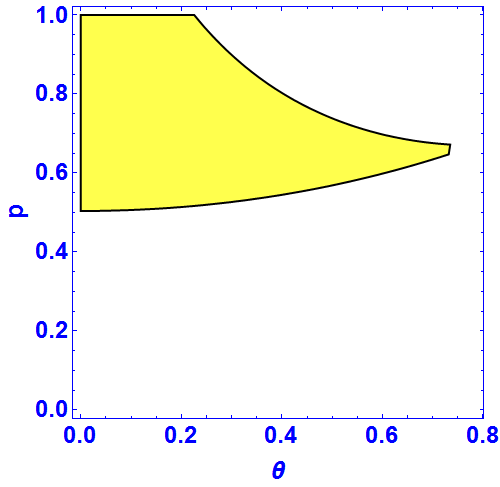}
	\caption{\emph{Shaded region gives the restrictions imposed on the state parameters for which enhancement of DIED is observed via the multiple swapping procedure under the restriction of $p_1\geq 0.670236$ over the state parameter $p_1$ of $\rho_2$. }}\label{pic3}
\end{figure}

Thus there exists a range of the state parameters $(p,p_1)$ where the initial states $\rho_i(i=1,2,3)$ do not violate the inequality Eq.(\ref{semi}), but after distributing them in the network and executing the protocol, final states $\rho_4$ violates it. The range of state parameters in which enhancement of detection of semi-device independent entanglement is observed by our protocol is given by: $p_1 \leq 0.670236$ and
\begin{equation}\label{r1}
\frac{2p \cos^2[\theta]}{1+p \cos[2 \theta]} < p < \frac{1.1831}{0.7324+1.0328 \sin[2\theta]}
\end{equation}
which indicates a clear advantage as shown in Fig.\ref{pic3}.

\section{CONCLUSION}
In a nutshell, our present topic of discussion may be considered as a  contribution in the field of device independent entanglement detection which minimizes the requirement of precise control over measurement devices by an experimentalist in an experimental detection of entanglement. More precisely, in our work we have shown that it is possible to enhance device independent detection of genuine tripartite entanglement in some suitable measurement context. For our purpose, we have considered four DIEWs given by Mermin\cite{Mer}, Uffink\cite{Uff}, Bancal\cite{Ban} and Liang et. al.\cite{Lia}, out if which the DIEW given by Bancal et.al.\cite{Ban} emerges to be the most efficient. We have designed a state preparation protocol (prior to receiving final measurements), particularly an entanglement swapping procedure involving six distant observers via which genuine tripartite entanglement of the resultant(swapped) state, generated by using three initial tripartite states(whose entanglement cannot be detected by the standard DIEWs), can be detected by these standard DIEWs(used for testing entanglement of the initial states) after performing the state preparation.\\

\textit{Acknowledgment}-We would like to gratefully acknowledge fruitful discussions with Prof. Guruprasad Kar. We also thank Tamal Guha and Mir Alimuddin for useful discussions. AM acknowledges support from the CSIR project 09/093(0148)/2012-EMR-I.

\appendix
\section{Condition for violation of DIEWs}\label{AppA}
We are now going to enlist the DIEWs which are used as tools for DIED in main text. To start with one can consider the device-independent-entanglement-witness provided by the Mermin polynomial \cite{Mer} as the simplest example for detecting genuine tripartite entanglement \cite{See}:
\begin{equation}\label{m}
M = |\langle A_1B_0C_0\rangle + \langle A_0B_1C_0\rangle + \langle A_0B_0C_1\rangle - \langle A_1B_1C_1 \rangle| \leq 2\sqrt{2}
\end{equation}
In \cite{Uff} Uffink, designed another nonlinear Bell-type inequality which has been extensively used for this purpose:\\
$$\langle A_1B_0C_0 +  A_0B_1C_0 + A_0B_0C_1 -  A_1B_1C_1 \rangle^2 + $$
\begin{equation}\label{m1}
\langle A_1B_1C_0 +  A_0B_1C_1 + A_1B_0C_1 -  A_0B_0C_0 \rangle^2 \leq 8
\end{equation}
In recent times, Bancal et al.\cite{Ban} have provided a more efficient $3$-settings Bell inequality(already discussed in details in main text) which can be used as a DIEW to detect genuine tripartite entanglement:\\
More than $3$ setting DIEWs are also provided in \cite{Pal}. However in our present topic of discussion, we restrict our search for only $3$-settings Bell inequalities due to obvious computational complexity. More recently another $2$ settings DIEW has been designed by Liang et al.\cite{Lia} :\\
$$\frac{1}{4}(\langle A_0B_0C_0\rangle + \langle A_0B_1C_0\rangle + \langle A_0B_0C_1\rangle + \langle A_0B_1C_1\rangle +$$
\begin{equation}\label{m4}
\langle A_1B_0C_0\rangle + \langle A_1B_1C_0\rangle +\langle A_1B_0C_1\rangle
- 3 \langle A_1B_1C_1\rangle) \leq \sqrt{2}.
\end{equation}
Now we present the detailed proofs of the results stated in the main text. To obtain the condition of violation of each of the DIEWs (Eqs.(\ref{m}, \ref{m1}, \ref{baneq}, \ref{m4})) in terms of state parameters for each of the initial states $\rho_i (i = 1, 2, 3)$ and final state $\rho_4$, we apply the same method as used in \cite{Kau}. First we find the condition of violation(in terms of state parameters) of the DIEW given in Eq.(\ref{m}) for the initial state $\rho_1$. We consider the following measurements: $A_0 = \vec{x}.\vec{\sigma_1} $ or $A_1 = \vec{\acute{x}}.\vec{\sigma_1}$ on $1^{st}$ qubit, $B_0 = \vec{y}.\vec{\sigma_2} $ or $B_1 = \vec{\acute{y}}.\vec{\sigma_2}$ on $2^{nd}$ qubit, and $C_0 = \vec{z}.\vec{\sigma_3} $ or $C_1 = \vec{\acute{z}}.\vec{\sigma_3}$ on $3^{rd}$ qubit, where $\vec{x},\vec{\acute{x}},\vec{y},\vec{\acute{y}}$ and $\vec{z},\vec{\acute{z}}$ are unit vectors and $\sigma_i$ are the spin projection operators that can be written in terms of the Pauli matrices. Representing the unit vectors in spherical coordinates, we have, $\vec{x} = (\sin\theta a_0 \cos\phi a_0, \sin\theta a_0 \sin\phi a_0, \cos\theta a_0), ~~\vec{y} = (\sin\alpha b_0 \cos\beta b_0, \sin\alpha b_0 \sin\beta b_0, \cos\alpha b_0) $ and $\vec{z} = (\sin\zeta c_0 \cos\eta c_0, \sin\zeta c_0 \sin\eta c_0, \cos\zeta c_0) $ and similarly, we define, $\vec{\acute{x}},\vec{\acute{y}}$ and $\vec{\acute{z}}$ by replacing $0$ in the indices by $1$. Then the value of the operator $M$ (Eq.(\ref{m})) with respect to the state $\rho_1$ (Eq.(\ref{1})) gives:
 \begin{widetext}
 $M(\rho_1) = -\cos\alpha b_1 (-1 + p + p \cos2 \theta)(\cos\zeta c_0\cos\theta a_1+\cos\zeta c_1\cos\theta a_0)-\sin\alpha b_1(p_1 \sin2\theta_1)(\cos(\beta b_1+\eta c_1+\phi a_0)\sin\zeta c_1\sin\theta a_0+\cos(\beta b_1+\eta c_0 +\phi a_1)\sin\zeta c_0\sin\theta a_1)+\cos\alpha b_0 (-1 + p + p \cos2 \theta)(\cos\zeta c_0\cos\theta a_0-\cos\zeta c_1\cos\theta a_1)$+
 \begin{equation}\label{A1}
   \sin\alpha b_0(p \sin2\theta)(\cos(\beta b_0+\eta c_0+\phi a_0)\sin\zeta c_0\sin\theta a_0-\cos(\beta b_0+\eta c_1+\phi a_1)\sin\zeta c_1\sin\theta a_1).
 \end{equation}
 \end{widetext}
Hence in order to get maximum value of $S(\rho_1)$, we have to perform maximization over $12$ measurement angles. Now if we maximize the last equation with respect to $\alpha b_0$ and $\alpha b_1$, we have
\begin{widetext}
$M(\rho_1) \leq \sqrt{((X)(\cos\zeta c_0\cos\theta a_1+\cos\zeta c_1\cos\theta a_0))^2+(Y)^2(A_{110}\sin\zeta c_1\sin\theta a_0 + A_{101}\sin\zeta c_0\sin\theta a_1)^2}$
\begin{equation}\label{A2}
+ \sqrt{((X)(\cos\zeta c_0\cos\theta a_0-\cos\zeta c_1\cos\theta a_1))^2 +(Y)^2(A_{000}\sin\zeta c_0\sin\theta a_0-A_{011}\sin\zeta c_1\sin\theta a_1)^2}
\end{equation}
\end{widetext}
Where $X = -1 + p + p \cos2 \theta $, $Y = p \sin2\theta$, and $A_{ijk} = \cos(\beta b_i+\eta c_j +\phi a_k) (i, j, k \in \{0,1\})$. The last inequality is obtained by using the inequality $x\cos\theta + y \sin\theta \leq \sqrt{x^2 + y^2}$. It is clear from the symmetry of the measurement angles $\theta a_0$ , $\zeta c_0$ and $\theta a_1$ , $\zeta c_1$ that the right hand side of Eq.(\ref{A2}) gives maximum value when $\theta a_0 = \zeta c_0$ and $\theta a_1 = \zeta c_1$. Hence Eq.(\ref{A2}) takes the form:
\begin{widetext}
$$M(\rho_1) \leq \sqrt{((X)(2\cos\theta a_0\cos\theta a_1))^2+(Y \sin\theta a_0\sin\theta a_1)^2(A_{110} + A_{101})^2}  +$$
\begin{equation}\label{A3}
 \sqrt{((X)(\cos^2\theta a_0-\cos^2\theta a_1))^2 +(Y)^2(A_{000}\sin^2\theta a_0-A_{011}\sin^2\theta a_1)^2}
\end{equation}
\end{widetext}
Again we maximize it with respect to $\theta a_1$. Critical point $0$ or $\frac{\pi}{2}$ gives the maximum value depending on values of the state parameters. For the critical point $0$, Eq.(\ref{A3}) becomes
\begin{equation}\label{A4}
M(\rho_1) \leq   \sqrt{(2 X \cos\theta a_0)^2} +  \sqrt{\sin^4\theta a_0(X^2+Y^2)}
\end{equation}
where we have chosen $A^2_{000} = 1$. Maximizing over $\theta a_{0}$, we get
\begin{equation}\label{A5}
M(\rho_1) \leq \frac{2 X^2 + Y^2}{\sqrt{X^2+Y^2}}
\end{equation}
the maximum being obtained for $\cos\theta a_0 = \frac{|X|}{\sqrt{X^2 + Y^2}}$. For the other critical point $\frac{\pi}{2}$, Eq.(\ref{A3}) takes the form:
$$M(\rho_1) \leq \sqrt{(Y \sin\theta a_0)^2(A_{110} + A_{101})^2}$$
$$+ \sqrt{X^2\cos^4\theta a_0 + Y^2(A_{000}\sin^2\theta a_0-A_{011})^2}$$
$$\leq \sqrt{4(Y \sin\theta a_0)^2} + \sqrt{X^2\cos^4\theta a_0 + Y^2(\sin^2\theta a_0+1)^2}$$
\begin{equation}\label{A6}
\leq 4 |Y|
\end{equation}
The second inequality in Eq.(\ref{A6}) is obtained from the first by setting $A_{110} = 1$, $A_{101} = 1$, $A_{000} = 1$ and $A_{011} = -1.$ The final inequality is achieved when $\theta a_0 = \frac{\pi}{2}$ . Two sets of measurement angles which realize the two values $\frac{2 X^2 + Y^2}{\sqrt{X^2+Y^2}}$(Eq.(\ref{A5})) and $4 |Y|$ (Eq.(\ref{A6})), are $\theta a_0 = \alpha b_0 = \zeta c_0 = \cos^{-1}(\frac{|X|}{\sqrt{X^2 + Y^2}}) $, $\theta a_1 = \alpha b_1 = \zeta c_1 = 0 $, $\beta b_i= \eta c_i = \phi a_i = 0$ (i = 0, 1) and $\theta a_i = \alpha b_i = \zeta c_i = \frac{\pi}{2}$(i = 0, 1) , $\beta b_0 = \eta c_0 = \phi a_0 = 0$, $\beta b_1 = -\eta c_1 = -\phi a_1 = \frac{\pi}{2}$ respectively. Hence from Eq.(\ref{A5}) and Eq.(\ref{A6}), we have
\begin{equation}\label{A7}
M(\rho_1) \leq \max[\frac{2 X^2 + Y^2}{\sqrt{X^2+Y^2}} , 4 |Y|] .
\end{equation}
Clearly, $\frac{2 X^2 + Y^2}{\sqrt{X^2+Y^2}}\leq 2 < 2 \sqrt{2} $ for any value of $p \in [0,1]$ and $0 \leq \theta \leq \frac{\pi}{4}.$ So the initial state $\rho_1$ violates the DIEW based on Mermin expression (Eq.(\ref{m})) if
\begin{equation}\label{A7}
4 |Y| = 4 | p | \sin2\theta > 2 \sqrt{2} .
\end{equation}
The last inequality is considered as the condition of violation of the DIEW based on Mermin expression for the initial state $\rho_1$. We have applied the same method over other states $\rho_i$ (i = 2, 3, 4)to find the condition of violation of the DIEW based on Mermin expression. For other
DIEWS (Eqs.(\ref{m1}), (\ref{baneq}), (\ref{m4})), we have made analysis in similar manner so as to obtain the condition of violation for each of states $\rho_i$. All the conditions are summarized in Table.\ref{table}.

However among the four DIEWs given by Mermin(Eq.(\ref{m})), Uffink(Eq.(\ref{m1})), Bancal et al.(Eq.(\ref{baneq})) and Liang et al.(Eq.(\ref{m4})), the one given by Bancal et al. turns out to be the most efficient for this purpose.
 The DIEW based on Bancal et al. polynomial (Eq.(\ref{baneq})) can thus detect genuine tripartite entanglement in a device-independent way
in $\rho_1$ for $p > \frac{2}{3\sin 2\theta}$(see Table\ref{table1}.). As $\frac{2}{3\sin 2\theta} < \frac{1}{\sqrt{2}\sin2\theta} < \frac{3\sqrt{2}}{5\sin2\theta}$, so the DIEW based on Bancal et al. polynomial (Eq.(\ref{baneq})) is the most efficient DIEW for the state $\rho_1$ to detect genuine tripartite entanglement among all the standard DIEWs considered in Eqs.((\ref{m}), (\ref{m1}), (\ref{baneq}), (\ref{m4})). Similarly by comparing the range of violation of $p_1$ (for the state $\rho_2$) and $p$ (for the state $\rho_3$, $\rho_4$), one can check that Bancal et al. Bell inequality is the best DIEW for the other states $\rho_i$ (i = 2, 3, 4) to detect genuine tripartite entanglement compared to other standard DIEWs.\\

\end{document}